\documentclass[pra, showpacs, twocolumn, floatfix]{revtex4}

\usepackage{graphicx}
\usepackage{amsmath, amsfonts, amssymb, bm}
\usepackage{nicefrac}

\usepackage{amstext}
\usepackage{mathrsfs}
\usepackage[]{psfrag}

\newcommand{\dirac}{\displaystyle{\not}}
\newcommand{\e}{\text{e}}
\psfragscanoff
\providecommand{\abs}[1]{\lvert#1\rvert}

\begin{document}
\title{Few--Photon Electron--Positron Pair Creation by Relativistic Muon Impact on Intense Laser Beams}
\author{Sarah J. M\"uller and Carsten M\"uller}
\affiliation{ Max-Planck-Institut f\"ur Kernphysik, 
Saupfercheckweg 1, D-69117 Heidelberg, Germany}

\begin{abstract}%
Electron--positron pair production in combined laser and Coulomb fields is studied. To this end, the Feynman diagram for multiphoton pair creation by muon impact on a circularly polarized high--frequency laser beam is evaluated within the framework of laser-dressed quantum electrodynamics employing relativistic Volkov states. In the limit of low laser intensity, the result is shown to coincide with the known expression for multiphoton pair creation by a proton which is treated as an external Coulomb field. A scaling of the total pair creation rate is analyzed. The recoil distribution is calculated numerically and its dependence on the projectile mass is discussed. Energy spectra of the created particles and angular spectra of the scattered muon are presented.
\end{abstract}


\maketitle

\section{Introduction}

The creation of matter--antimatter particle pairs in external electromagnetic fields is a characteristic effect in relativistic quantum theory. 
For example, a very strong static electric field may cause the spontaneous creation of $e^+e^-$ pairs, if the field strength surmounts the critical value of $E_{cr}=m_e^2c^3/e\hbar=1.3\times 10^{16}$V/cm found already by \textsc{Sauter} \cite{Sauter}. Here, $m_e$ denotes the electron mass, $e$ the elementary charge unit, $c$ the light velocity in vacuum, and $\hbar$ Planck's constant. In recent years, interest in nonlinear pair creation processes via multiphoton absorption from external laser fields has raised \cite{Marklund,Salamin06,EhlotzkyRev}. Yet a periodically changing electromagnetic plane--wave field, like a laser beam, cannot by itself lead to pair creation, independently of its frequency or intensity as shown by \textsc{Schwinger} \cite{Schwinger}. Thus, an additional source is needed such as e.~g. a second laser beam \cite{Brezin70,Popov71,Schuetzhold08,Rufus,Cheng}, a single (non--laser) photon \cite{Reiss62,Reiss71,Narozhnyi,Bell}, or a charged particle. Here, we consider $e^+e^-$ pair production by the collision of charged particles with laser fields. \\
There are two possible channels via which the creation of the pairs can take place. The first channel is of Bethe--Heitler type where the pair is created by a virtual photon from the Coulomb field of the projectile particle and $r$ real photons from the laser field. While \textsc{Bethe} and \textsc{Heitler} originally treated the linear case of pair production by a single high--energy photon ($r=1$) \cite{BetheHeitler}, below we focus on the nonlinear Bethe--Heitler effect which involves the absorption of $r>1$ laser photons and thus depends on the laser intensity in a nonlinear way. In the case of a muon projectile, the symbolic equation for this process is $\mu+r\omega\rightarrow\mu+e^+e^-$, with the laser frequency $\omega$. In the second possible channel, the pair creation takes place by the collision of the laser beam with a real photon (nonlinear Breit--Wheeler process \cite{BreitWheeler}), with the latter stemming from a Compton scattering event, for instance. This indirect mechanism is important for light projectile particles, such as electrons.
\\
In 1997, laser--induced $e^+e^-$ pair creation has been experimentally observed at SLAC (Stanford, USA). In this pioneering experiment, highly relativistic electrons with an initial energy of $46$GeV were scattered by an optical laser pulse with an intensity of $10^{18}$W/cm$^2$ \cite{SLAC, SLAC2}. The electrons were Compton back--scattered, thus emitting $30$GeV $\gamma$--photons which in turn collided with the laser beam. In this collision, $r=5$ laser photons of $\hbar\omega\approx 2$eV combined their energies with the $\gamma$--photon and thus produced an $e^+e^-$ pair via the nonlinear Breit--Wheeler process $\omega_C+r\omega\rightarrow e^+e^-$ \cite{Reiss62,Reiss71,Narozhnyi,Bell}.
\\Inspired by the SLAC experiment, several theoretical studies bear on the---experimentally not yet observed---Bethe--Heitler creation of electron--positron pairs in the collision of a relativistic proton or nucleus with a laser beam \cite{Yakovlev,Mittleman87,Roshchupkin9601,Dietz98,Dremin02,Avetissian03,Muller03a,Carsten,Kaminski06,Milstein06,Kuchiev07,CarlusCarsten}. In the rest frame of the projectile particle, the energy of the laser photons is significantly higher than in the laboratory frame because of the relativistic Doppler shift. 
Recently, $e^+e^-$ pair creation via the collision of a high--energy neutrino with an intense laser beam was calculated by \textsc{Tinsley} \cite{Tinsley05}. There, the creation of the $e^+e^-$ pair takes place via an intermediate $Z^0$ boson.
\\The field--induced pair creation process can be divided into three regimes by means of the so--called laser intensity parameter 
\begin{equation}\label{xifirstdef}
\xi=\frac{e
a}{m_ec^2}\,,
\end{equation}
$a$ being the amplitude of the laser potential. 
One distinguishes the {\em tunnel regime}, where $\xi\gg 1$ (and $a\omega/c\ll E_{cr}$) and the pair creation is similar to a tunnelling process from the negative- to the positive--energy continuum, the {\em above--threshold regime} similar to the above--threshold ionization of atoms or molecules in strong laser fields where $\xi\sim1$, and the {\em multiphoton regime} with $\xi\ll 1$. 
The typical photon orders yielding the main contribution to the total pair production rate in the tunnel and above--threshold regime are much larger than the minimum order $r_0$ which is required by energy--momentum conservation (see Eq. \eqref{threshold} below). In the multiphoton regime, the main contribution to the pair production process comes from the minimum number $r_0$ allowing the creation to take place. In this regime, the total pair creation rate $R$ scales with $R\propto\xi^{2r_0}$.
\\
\begin{figure}
 \includegraphics[width=0.3\textwidth]{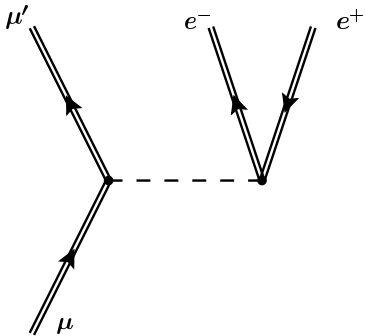}
\caption{Feynman diagram of the considered process in the lowest order in $\alpha_f$. The dashed line represents the virtual photon propagated between the projectile and the electron--positron vertices, and the double lines stand for the exact lepton wave functions in the laser field (Volkov states).}\label{FeynDiag1}
\end{figure}\begin{figure}
 \includegraphics[width=0.49\textwidth]{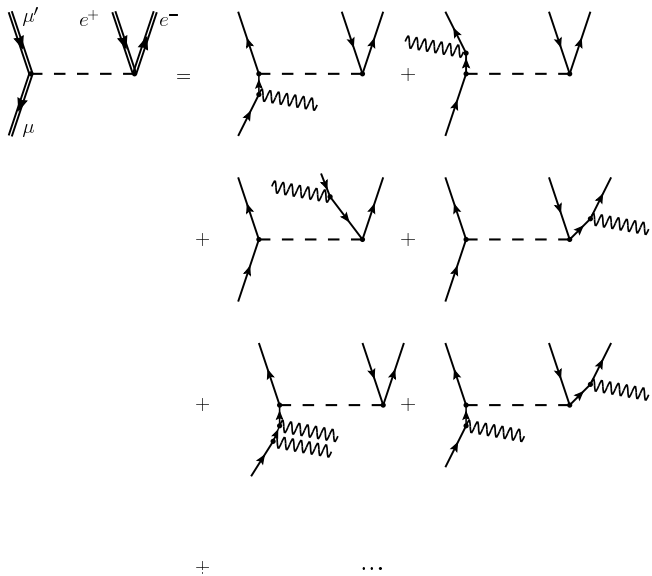}
\caption{Decomposition of the Feynman graph 
in Fig. \ref{FeynDiag1} into a perturbation series. The left--hand side represents the total Feynman diagram, where the leptons are represented by exact lepton wave functions in the laser field, i.e. the Volkov states (double lines) (Eq. \eqref{VolkovElectron}). On the right--hand side, we show a few examples of which underlying graphs constitute the left--hand side Feynman diagram. The wiggled lines represent real laser photons absorbed in the pair creation process. The first two rows on the right--hand side show the four leading graphs for one--photon processes, whereas the third row shows two exemplary two--photon processes. Higher orders involve both absorption and emission of laser photons, where the net absorbed energy must surmount the energy threshold (Eq. \eqref{threshold}).}\label{FeynDiagExpansion1}
\end{figure}
 %
%

In the present paper, we study Bethe--Heitler $e^+e^-$ creation in the collision of a relativistic muon with a high--frequency circularly polarized laser beam within the multiphoton regime. In comparison with already existing calculations for projectile nuclei, which are treated as infinitely heavy particles providing an external Coulomb field, the present approach has two advantages: 
firstly, the fact that the considered projectile particles are leptons and as such fundamental particles, allows us to treat the process exactly in terms of (laser--dressed) quantum electrodynamics up to the leading order in the coupling constant $\alpha_f$ 
(Fig. \ref{FeynDiag1}). 
Secondly, the treatment of the projectile nuclei as external fields does not account for their finite mass and thus neglects any recoil effects on the projectile. The approach pursued in this paper allows to study the said recoil effects and their dependence on the projectile mass and the photon order. We show in particular that the muon recoil becomes the more pronounced the more laser photons (of equal total energy) participate in the pair production.
\\
%
The laser--dressed Feynman graph in Fig. \ref{FeynDiag1} can be expanded into different orders of absorbed laser photons (Fig. \ref{FeynDiagExpansion1}). The pair creation process consists of all possible $r$--photon processes, which themselves are composed of all possible processes involving combinations of absorbed or emitted laser photons at each of the two vertices leading to $r$ net absorbed photons. 
%
Note that, since we study low laser intensities, any of these diagrams could be calculated by means of ordinary quantum electrodynamics (QED) within $r$--th order perturbation theory in the photon field. However, the framework of laser--dressed QED employing relativistic Volkov states is more convenient for our purposes \cite{Reiss62, Reiss71, Narozhnyi, ldQED}. 
%
%
%
%
%
\\In the rest frame of the incoming projectile, the photon energy is higher than in the laboratory system by the Doppler factor $\sqrt{(1+\beta)/(1-\beta)}$, where $\beta$ is the ratio of the muon's lab--frame velocity to the light velocity. Due to the finite mass of the muon projectile, not only the energy $2m_ec^2$ for the creation of an electron and a positron must be surmounted, but also the recoil energy of the scattered projectile must be provided. By calculating the Mandelstam variable for the invariant mass of the process, one finds the threshold relation in the rest frame of the incoming muon for $r$ absorbed laser photons of energy $\hbar \omega$,
\begin{equation}\label{threshold}
 r\cdot\hbar\omega\ge2m_ec^2\bigl(1+\frac{m_e}{M}\bigr)\,,
\end{equation}
$M$ being the mass of the projectile. The muon rest mass is $M=105.6$MeV/c$^2$, i.~e. about 200 times the rest mass of an electron. Since the probability for Compton scattering is inversely proportional to the scattering particle's mass, the Compton channel occuring in the SLAC experiment will be highly suppressed for muon projectiles. 
\\We note that the process ${\mu+r\omega\rightarrow\mu+e^+e^-}$ is formally related to some other processes via crossing symmetry: for instance, electron--muon scattering in strong laser fields ${e^-\mu^-\rightarrow e^-\mu^-}$ \cite{Nedoreshta}, or muon pair creation from positronium in laser fields ${e^+e^-\rightarrow\mu^+\mu^-}$ \cite{CarstenMuonpairs} (see also \cite{Thoma09,KuchievMuonpairs,NedoPairs}). Moreover, there are certain similarities to laser--assisted M\o ller scattering ${e^-e^-\rightarrow e^-e^-}$ \cite{Panek} and laser--assisted Bhabha scattering ${e^+e^-\rightarrow e^+e^-}$ \cite{Denisenko}. Also, a laser--assisted Bethe--Heitler process, where a high--frequency $\gamma$--photon and a nucleus collide within a background laser field, ${Z+\gamma\rightarrow Z+e^+e^-}$, has been discussed \cite{Erik}. Besides the creation of matter--antimatter particle pairs, there are also other interesting QED effects that occur in combined laser and Coulomb fields, such as photon fusion \cite{Toni1} or Delbr\"uck scattering \cite{Toni2}. Efficient pair production by the linear Bethe--Heitler effect through bremsstrahlung photons in a laser--induced plasma has been observed recently \cite{Beiersdorfer}.\\

The paper is organized as follows. In Section \ref{TheorySec}, we derive an expression for the fully differential pair production rate. In Section \ref{Resultsec}, the results of the numerical calculation of pair creation processes involving the net absorption of $r=1,2$ or $3$ photons are given, including total and differential pair creation rates. Section \ref{ConclusionSec} gives a summary of the paper. In the Appendix \ref{compApp}, we summarize the calculation for proton projectiles and compare it to our calculation for muon projectiles. 
\\We employ a natural units system in which $\hbar=c=1$ and $e=\sqrt{\alpha_f}$ ($\alpha_f=1/137$ being the fine structure constant). We make use of the metric tensor $g^{\mu\nu}=$diag$(+---)$, so that the scalar product of two four--vectors $p^\mu=(p^0,\bm{p})$ and $q^\mu=(q^0,\bm q)$ reads $(pq)=p^0q^0-\bm{pq}$. Furthermore, we employ Feynman slash notation for four--products of four--vectors with the Dirac matrices $\gamma^\mu$, $\dirac p=(\gamma p)$.

\section{Theory}\label{TheorySec}
The Feynman graph corresponding to the process under consideration is shown in Fig. \ref{FeynDiag1}. 
In the collision with the laser beam, the muon experiences a certain recoil and emits a virtual photon, which decays into an electron--positron pair. 
The amplitude of the process then reads
\begin{multline}
 \mathscr{S} = \frac{\alpha_f}{i} \iint d^4x d^4y \overline{\psi}_{p_- s_-}(x) \gamma^\mu \psi_{p_+ s_+}(x) \\ \times \mathcal{D}_{\mu \nu}(x-y) \overline{\Psi}_{P' S'}(y) \gamma^\nu \Psi_{P S}(y)\,.\label{Ampli}
\end{multline}
Here, $x$ and $y$ denote the space--time coordinates of the produced pair and the scattering particle, respectively. $\mathcal{D}_{\mu \nu}(x-y)$ is the propagator for the virtual photon propagating between the two vertices. The muon vertex is described by $\overline{\Psi}_{P' S'}(y) \gamma^\nu \Psi_{P S}(y)$, and the electron--positron vertex by $\overline{\psi}_{p_- s_-}(x) \gamma^\mu \psi_{p_+ s_+}(x)$ with the Dirac matrices $\gamma^\mu$. $\psi$ and $\Psi$ are the Volkov states \cite{Volkov,LandauLifschitz} of the respective particles which solve the Dirac equation for spin--$\nicefrac{1}{2}$ particles in an electromagnetic field,
\begin{equation}\label{DiracEq}
 (\dirac \partial + e\dirac A-m)\psi=0\,,
\end{equation}
with the laser potential $A^\mu=(0,\bm{A})$ in the radiation gauge. We consider a circularly polarized laser field, so that the four potential reads $A^\mu(\eta)=a_1\cos(\eta)+a_2\sin(\eta)$ with the laser phase $\eta:=kx$, where $k$ is the wave vector of the laser field. The four--vectors $a_{1/2}$ read $a_{1}=(0,a,0,0)$ and $a_2=(0,0,a,0)$, $a$ being the amplitude of the laser potential. The state vector for an electron with the kinetic momentum $p_-$ and the spin projection $s_-$ thus can be written as
\begin{multline}\label{VolkovElectron}
 \psi_{p_- s_-}(x)=\sqrt{\frac{m}{Vq_-^0}} \left(1 - \frac{e\displaystyle{\not} k  \displaystyle{\not}A }{2 (k p_-)} \right) u_{p_- s_-} \e^{i\mathcal S}\,,
\end{multline}
with the action
\begin{equation}\label{action}
 \mathcal S=-(q_- x) + \frac{e(a_1p_-)}{ (k p_-)}  \sin(\eta) - \frac{e( a_2p_-)}{(kp_-)} \cos(\eta) \,.
\end{equation}
Here, $m=m_e$ is the electron mass, $V$ a normalization volume, and $u_{p_- s_-}$ is a free Dirac spinor \cite{BjorkenDrell}. $q_-$ is the effective momentum of the electron in the laser field \cite{LandauLifschitz}, 
\begin{equation}
 q^\mu_-=p^\mu_-+\frac{e^2a^2}{2(kp_-)}k^\mu.
\end{equation}
The corresponding effective mass is $m_*=m(1+\xi^2)$ with the laser intensity parameter $\xi$ of Eq. \eqref{xifirstdef}. The Volkov states for the positron can be obtained from \eqref{VolkovElectron} and \eqref{action} by replacing $p^\mu$ by $-p^\mu$ and $u_{p_- s_-}$ by a corresponding antiparticle spinor $v_{p_+ s_+}$. Replacement of the coordinate $\eta=kx$ by $\kappa=ky$, the mass $m$ by the projectile mass $M$, the kinetical momentum $p_-$ by $P$ or $P'$ for the incoming or scattered muon, respectively, and the effective momentum $q_-$ by $Q$ or $Q'$ yields the Volkov states for the initial and scattered muon. The effective muon mass is obtained by $M_*=M(1+\Xi^2)$, and the corresponding intensity parameter reads $\Xi=ea/M=\xi m/M$. 
\\We concentrate our study on the multiphoton regime $\xi\ll 1$, where the intensity of the laser field is comparatively low and the photon energy is high. Because of the low intensity, the electric field is much smaller than the critical value $E_{cr}=1.3\times10^{16}$V/cm \cite{Sauter}. Therefore, we neglect vacuum polarization effects which become important only at near--critical laser intensities $I\gtrsim10^{29}$W/cm$^2$ \cite{Becker75, ErikPropagatorFootn,ErikPropagator} and employ a free photon propagator for the description of the virtual photon which is propagated between the two vertices:
\begin{equation}\label{Photonprop}
 \mathcal{D}_{\mu\nu}(x-y) = \int \frac{d^4q}{(2\pi)^4} \frac{4 \pi \e^{i q (x-y)}}{q^2} g^{\mu\nu}\,,
\end{equation}
with the integration variable $q$ describing the momentum of the virtual photon.
\\It is important to note that the amplitude in Eq. \eqref{Ampli} fully accounts for the interaction of the leptons with the laser field by employing the exact solutions to the Dirac equation, i.~e. the Volkov states; the interaction between the leptons and the QED vacuum however is taken into account only to lowest order in $\alpha_f$. A similar approach can be found in \cite{CarstenMuonpairs,NedoPairs,Panek,Denisenko}.\\

The space--time integrals in \eqref{Ampli} can be performed by expanding the individual vertex expressions into Fourier series. Using the generating function of the regular cylindrical Bessel functions \cite{Abramowitz}, one finds  
\begin{widetext}
\begin{multline}\label{transAmpl}
  \mathscr{S}
= \frac{\alpha_f}{i} \frac{2(2\pi)^5Mm}{V^2} \int \,\frac{d^4q}{q^2\sqrt{q_+^0q_-^0Q^{'0}Q^0}}\sum_{n,N}\mathcal{M}^\mu(e^+e^-|n)\mathcal{M}_\mu(\mu,\mu'|N)
\delta(q+q_++q_--nk)\delta(q+Q+Nk-Q')\,.
 \end{multline}
\end{widetext}
The integer numbers $n$ and $N$ correspond to the numbers of absorbed laser photons at the electron--positron and projectile vertex, respectively. Note that $n$ and $N$ may become negative, amounting to photon emission at the respective vertex, whereas the total number of absorbed photons, $r=n+N$, may not.
The electronic spinor--matrix product $\mathcal{M}^\mu(e^+e^-|n)$ 
can be written as
\begin{align}\label{sp-mat-pr}
  \mathcal{M}&^\mu(e^+,e^-|n)
:=\bar{u}_{p_-s_-}\biggl(\bigl(\gamma^\mu - \frac{e^2a^2k^\mu}{2(kp_+)(kp_-)}\dirac k \bigr)\cdot B_n \nonumber\\
&+ \frac{e}{2}\Bigl(\bigl[\frac{1}{(kp_+)}\gamma^\mu\dirac k \dirac a_1 - \frac{1}{(kp_-)}\dirac a_1 \dirac k\gamma^\mu \bigr]\cdot C_n \nonumber\\
&+ \bigl[\frac{1}{(kp_+)}\gamma^\mu\dirac k \dirac a_2 - \frac{1}{(kp_-)}\dirac a_2 \dirac k\gamma^\mu \bigr]\cdot D_n \Bigr)\biggr)v_{p_+ s_+}\,,
 \end{align}
with the coefficients
\begin{align}\label{coefficients}
 B_n&=J_n(\bar{\alpha})\e^{in\eta_0}\,,\nonumber\\
  C_n&=\frac{1}{2}\Bigl(J_{n+1}(\bar{\alpha})\,\e^{i(n+1)\eta_0}+J_{n-1}(\bar{\alpha})\,\e^{i(n-1)\eta_0} \Bigr)\,,\nonumber\\
  D_n&=\frac{1}{2i}\Bigl(J_{n+1}(\bar{\alpha})\,\e^{i(n+1)\eta_0}-J_{n-1}(\bar{\alpha})\,\e^{i(n-1)\eta_0} \Bigr)\,.
\end{align}
The functions $J_i(\bar\alpha)$ are the regular cylindrical Bessel functions of integer order. Their argument is $\bar\alpha=\sqrt{\alpha_1^2+\alpha_2^2}$ and the angle $\eta_0$ is given by $\cos\eta_0=\alpha_1/\alpha,$ $\sin\eta_0=\alpha_2/\alpha$ with
\begin{equation}\label{alphadef}
 \alpha_{j} = \frac{e(a_jp_-)}{(kp_-)}-\frac{e(a_jp_+)}{(kp_+)}\,,\quad j=1,2\,.
\end{equation}
Note that the Bessel functions account for the fact that in the multiphoton regime, the main contribution to the total pair production rate comes from the minimum number of absorbed photons $r_0$, for which the threshold relation \eqref{threshold} is fulfilled: for small arguments $\bar\alpha\ll1$, the Bessel functions scale with $J_n(\bar\alpha)\sim\bar\alpha^{\abs{n}}$ \cite{Abramowitz}. Since the laser field parameter $\xi$ is supposed to be small and it can be factored out in the argument of the Bessel functions, higher photon orders yield partial rates which are several orders of magnitude smaller than the one for $r_0$.\\
The corresponding muonic spinor--matrix product $\mathcal{M}_\mu(\mu,\mu'|N)$ is
\begin{align}\label{sp-mat-pr-muons}
  \mathcal{M}&^\nu(\mu,\mu'|N):=\bar{U}_{P'S'}\Biggl( \bigl(\gamma^\nu + \frac{e^2a^2k^\nu}{2(kP')(kP)}\dirac k \bigr)\cdot F_N\nonumber\\
&- \frac{e}{2}\biggl(\bigl[\frac{1}{(kP)}\gamma^\nu\dirac k \dirac a_1 + \frac{1}{(kP')}\dirac a_1 \dirac k\gamma^\nu \bigr]\cdot G_N \nonumber\\
&+ \bigl[\frac{1}{(kP)}\gamma^\nu\dirac k \dirac a_2 + \frac{1}{(kP')}\dirac a_2 \dirac k\gamma^\nu \bigr]\cdot H_N \biggr)\Biggr)U_{P S}\,.
 \end{align}
The different signs in Eq. \eqref{sp-mat-pr} and \eqref{sp-mat-pr-muons} occur because the former spinor--matrix product describes a particle and an antiparticle, whereas in the latter only one type of particle appears. The argument of the Bessel functions $J_N(\bar\beta)$ for this vertex reads
\begin{equation}\label{betadef}
 \bar{\beta}=\sqrt{\beta_1^2+\beta_2^2}\,,
\end{equation}
with 
\begin{equation}
\beta_j:=\frac{e(a_jP')}{(kP')}-\frac{e(a_jP)}{(kP)}\,,\quad j=1,\,2\,.
\end{equation}
Making use of the angle $\kappa_0$ defined by $\cos\kappa_0=\beta_1/\bar\beta,$ $\sin\kappa_0=\beta_2/\bar\beta$, the coefficients $F_N,$ $G_N$, and $H_N$ are defined similarly to Eq. \eqref{coefficients} involving $J_N(\bar\beta)$.
\\We can separate the amplitude into a sum of partial amplitudes $\mathscr{S}^{(r)}$ for one particular photon order $r=n+N$,
\begin{multline}\label{partAmpli}
\mathscr{S}^{(r)}
= \mathcal N' 
\sum_n \mathcal{M}^\mu(e^+e^-|n)\cdot\mathcal{M}_\mu(\mu,\mu'|r-n)\\
\cdot\frac{\delta(q_++q_-+Q'-Q-rk)}{(q+q_++q_--nk)^2}\,,
\end{multline}
with the factor 
\begin{equation}
 \mathcal N'=\frac{\alpha_f}{i}\cdot \frac{2(2\pi)^5Mm}{V^2\sqrt{q_+^0q_-^0Q^{'0}Q^0}}\,.
\end{equation}
Then, the total transition amplitude and its square read 
\begin{equation}\label{partialsq}
 \mathscr S=\sum_{r\ge r_0}\mathscr S^{(r)}\,, \quad\abs{\mathscr S}^2=\sum_{r\ge r_0}\abs{\mathscr S^{(r)}}^2\,,
\end{equation}
where $r_0$ is the minimal number for which the threshold relation \eqref{threshold} is fulfilled. Because of the delta function in Eq. \eqref{partAmpli}, there is no double sum over $r,r'$ so that the summation in \eqref{partialsq} can be pulled out. The square of the partial amplitude $\mathscr S^{(r)}$ is
\begin{widetext}
\begin{align}\label{squaredPartAmp}
 \abs{\mathscr{S}^{(r)}}^2
&=\abs{\mathcal N'}^2\sum_{n,n'}\frac{\mathcal{M}^\mu(e^+e^-|n)\mathcal{M}_\mu(\mu,\mu'|r-n) \mathcal{M}^\dagger_\nu(\mu,\mu'|r-n')\mathcal{M}^{\dagger\nu}(e^+e^-|n')TV\delta(q_++q_-+Q'-Q-rk)}{(2\pi)^4(q_++q_--nk)^2\cdot(q_++q_--n'k)^2}\,.
\end{align}
\end{widetext}
The factor $TV/(2\pi^4)$ stems from the square of the delta function \cite{BjorkenDrell}, leaving a time factor $T$ and a volume factor $V$ which is the same as in the normalization of the Volkov states \eqref{VolkovElectron}.
\\The total pair production rate is obtained from \eqref{squaredPartAmp} by averaging over the possible initial spin states, summing over the final spin states, integrating over the final momenta, and dividing the result by a unit time $T$:
\begin{equation}\label{totRate}
 R=\int\frac{Vd^3Q'}{(2\pi)^3}\int\frac{Vd^3q_+}{(2\pi)^3}\int\frac{Vd^3q_-}{(2\pi)^3}\frac{1}{2}\sum_{S}\sum_{S',s_+,s_-}\frac{\abs{\mathscr S}^2}{T}\,.
\end{equation}
With the completeness relation for the components of the Dirac spinors $u_{p_-s_-}$ and $v_{p_+s_+}$ \cite{BjorkenDrell}, we can perform the spin summation by introducing the matrices
\begin{align}\label{DeltaDef}
 {_r\Delta_\mu^n}=\Bigl[&\Bigl(\gamma_\mu + \frac{e^2a^2k_\mu\dirac k}{2(kP)(kP')} \Bigr)F_{r-n} \nonumber\\
&- \frac{e}{2}\Bigl(\frac{1}{kP}\gamma_\mu\dirac k \dirac a_1 + \frac{1}{kP'}\dirac a_1\dirac k\gamma_\mu\Bigr)G_{r-n} \nonumber\\
&- \frac{e}{2}\Bigl(\frac{1}{kP}\gamma_\mu\dirac k \dirac a_2 + \frac{1}{kP'}\dirac a_2\dirac k\gamma_\mu\Bigr)H_{r-n}\Bigr]
\end{align}
for the projectile vertex and 
\begin{align}\label{GammaDef}
 {_r\Gamma^\mu_n}=\Bigl[&\Bigl(\gamma^\mu - \frac{e^2a^2k^\mu\dirac k}{2(kp_+)(kp_-)} \Bigr)B_{n} \nonumber\\
 &+ \frac{e}{2}\Bigl(\frac{1}{kp_+}\gamma^\mu\dirac k \dirac a_1 - \frac{1}{kp_-}\dirac a_1\dirac k\gamma^\mu\Bigr)C_{n} \nonumber\\
&+ \frac{e}{2}\Bigl(\frac{1}{kp_+}\gamma^\mu\dirac k \dirac a_2 - \frac{1}{kp_-}\dirac a_2\dirac k\gamma^\mu\Bigr)D_{n}\Bigr]
\end{align}
for the vertex of the produced pair. We find
\begin{widetext}
\begin{multline}\label{SpinSum}
 \sum_{S,S',s_+,s_-}\mathcal{M}^\mu(e^+e^-|n)\mathcal{M}_\mu(\mu,\mu'|r-n) \mathcal{M}^\dagger_\nu(\mu,\mu'|r-n')\mathcal{M}^{\dagger\nu}(e^+e^-|n')\\
=\text{Tr}\Bigl({_r\Gamma^\mu_n}\frac{\dirac p_+-m}{2m}{_r\bar{\Gamma}^\nu_{n'}}\frac{\dirac p_-+m}{2m}\Bigr)\cdot\text{Tr}\Bigl({_r\Delta_\mu^n} \frac{\dirac P+M}{2M}{_r\bar{\Delta}_\nu^{n'}}\frac{\dirac P'+M}{2M} \Bigr)=:T^{nn'}_r\,.
\end{multline}
\end{widetext}
This trace product can be evaluated by the standard procedure.
\\Since the kinematics of this process are rather complicated in the rest frame of the incoming muon, we now transform into the center--of--mass (c.m.) system in order to perform the integrations in Eq. \eqref{totRate}. We can separate the total rate into partial rates $R=\sum_rR_r$ corresponding to the contributions of particular photon orders $r$ like we did for the transition amplitude. For each photon order $r$ there is a corresponding c.m. system. Let $\omega$ be the laser frequency in the rest frame of the incoming projectile muon and $k=\omega(1,0,0,1)$ the corresponding wave vector. In the c.m. system, the incoming muon moves along the $z$--direction and
\begin{equation}\label{cm1}
 \text{\textbf{Q}}_{cm, r} = \begin{pmatrix} Q_{cm,r}^1\\ Q_{cm,r}^2\\Q_{cm,r}^3\end{pmatrix}=-r\cdot\text{\textbf{k}}_{cm,r} = -r\begin{pmatrix} 0\\0\\ \omega_{cm,r}\end{pmatrix} \,,
\end{equation}
where the index ($cm,r$) indicates the c.m. system corresponding to the photon order $r$. The laser frequency transforms as
\begin{equation}
 r\omega_{cm,r}=r\omega\sqrt{\frac{1-\beta_{cm,r}}{1+\beta_{cm,r}}}=r\omega\sqrt{\frac{M_*}{M_*+2r\omega}}\,
\end{equation}
with the velocity 
\begin{equation}
 \beta_{cm,r}=\frac{\abs{\text{\textbf{Q}}_{cm,r}}}{Q^0_{cm,r}}=\frac{r\omega}{M_*+r\omega}\,.
\end{equation}
According to this, the center of mass as viewed from the projectile rest frame moves with the Lorentz factor ${\gamma_r=(1-\beta_{cm,r}^2)^{-1/2}}$. We can perfom one of the momentum integrations in Eq. \eqref{totRate} by making use of the three--dimensional $\delta$--function. This gives e.~g. ${(\text{\textbf{q}}_+)_{cm,r}=-\text{\textbf{Q}}'_{cm,r}-(\text{\textbf{q}}_-)_{cm,r}}$. 
\begin{figure}%
 \includegraphics[width=0.25\textwidth]{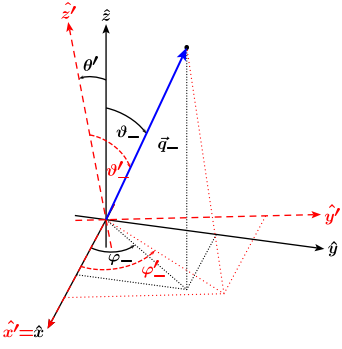}\caption{(Color online.) Visualization of the coordinate transformation which Eq. \eqref{DiffPartRate6} is based upon.}\label{TrafoPic}%
\end{figure}%
For the sake of brevity, we will omit the index $(cm,r)$ where there is no ambiguity. From now on, all quantities refer to the c.m. system.
\\The remaining integrations are performed in spherical coordinates. Because of the spherical symmetry we may choose the $x$--axis of the coordinate system freely. Thus, we may set the azimuth angle of the scattered muon ${\phi'=0}$ so that the integration yields the constant factor $2\pi$.
\\Regarding the remaining integrations, we perform a coordinate transformation following \textsc{Mork} \cite{Mork}. We regard the solid angle element of the created electron within a coordinate system where the $z'$--direction is given by the momentum vector of the scattered muon (see Fig. \ref{TrafoPic}). For this transformation, the Jacobi determinant is unity, $J\equiv1$, so that the partial rate in the c.m. system reads
\begin{align}\label{DiffPartRate6}
R^r_{cm,r}&= \frac{\alpha_f^2m^2M^2}{2\pi^2}\int\Bigl(\frac{\abs{\text{\textbf{q}}_-}\abs{\text{\textbf{Q}}'}}{ q_+^0Q^0} \cdot\sum_{n,n'}T^{nn'}_r\nonumber\\
&\cdot\frac{\delta(E - q_+^0-q_-^0-Q'^0)}{(q_++q_--nk)^2(q_++q_--n'k)^2}\nonumber\\
&\cdot \left. d(\cos\vartheta'_-)d\varphi'_-d(\cos\theta')dq_-^0dQ'^0\Bigr)\right|_{\text{\textbf{q}}_+= - \text{\textbf{Q}}' - \text{\textbf{q}}_-}\,,
\end{align}
with the abbreviation $E:=Q^0+rk^0$. The new polar angle of the electron, $\vartheta'_-$, is the angle between the momentum vectors of the electron and scattered muon and is found in the energy--conserving $\delta$--function (because $\text{\textbf{q}}_+= - \text{\textbf{Q}}' - \text{\textbf{q}}_-$). The argument of the $\delta$--function has one root at
\begin{equation}
 \cos\vartheta'^0_-=\frac{\left(E-Q'^0\right)^2-\abs{\text{\textbf{Q}}'}^2-2q_-^0\left(E-Q'^0\right)}{2\abs{\text{\textbf{Q}}'}\abs{\text{\textbf{q}}_-}}\,
\end{equation}
and from the requirement that $\abs{\cos\vartheta'^0_-}\le1$ follow the integration limits $\varepsilon^\pm$ for the energy $q^0_-$ of the created electron,
\begin{align}
 \varepsilon^\pm = \frac{1}{2}\left(E-Q'^0\pm \abs{\text{\textbf{Q}}'}\sqrt{1-\frac{2m_*^2}{E\left(Q^0-Q'^0\right)}}\right)\,.
\end{align}
Since these limits must be real numbers, one finds the upper limit for the integration over the energy $Q'^0$ of the scattered muon,
\begin{equation}\label{Qprime_max}
Q'^0\le \frac{m_*^2}{M_*^2}\left(2\abs{\text{\textbf{Q}}}-Q^0\left(2-\frac{M_*^2}{m_*^2}\right)\right)=:Q'^0_{\text{max}} \,,
\end{equation}
and the lower limit is $M_*$.
\\Thus we can write the partial rate in the c.m. system
\begin{align}\label{IntegrationCM}
& R_{cm,r}=\int_{M_*}^{Q'^0_{\text{max}}}dQ'^0\int_{\varepsilon^-}^{\varepsilon^+}dq_-^0\int_{-1}^{+1}d\cos\theta'\int_{0}^{2\pi}d\varphi'_- \mathcal N \nonumber\\
\cdot&\Bigl( \sum_{n,n'}\frac{T^{nn'}}{(q_++q_--nk)^2(q_++q_--n'k)^2} \left. \Bigr)\right|_{\text{\textbf{q}}_+= - \text{\textbf{Q}}' - \text{\textbf{q}}_-}^{\cos\vartheta'_-=\cos\vartheta'^0_-}\,,
\end{align}
with the factor
\begin{equation}
 \mathcal N=\frac{\alpha_f^2m^2M^2\abs{\text{\textbf{q}}_-}\abs{\text{\textbf{Q}}'}}{2\pi^2\cdot q_+^0Q^0}\,.
\end{equation}
%
Eq. \eqref{IntegrationCM} gives an expression for the partial rate corresponding to one particular photon order $r$ in the respective c.m. system. In order to obtain the total pair creation rate, we have to sum over all relevant orders in the rest frame of the initial projectile:
\begin{equation}\label{rate}
 R = \sum_{r=r_0}^\infty R_r = \sum_{r=r_0}^\infty  \gamma_r R_{cm,r} \,,
\end{equation}
where the Lorentz factor $\gamma_r$ accounts for the time dilation between the frames. In the multiphoton regime considered in the following, all the summands with $r>r_0$ yield negligibly small contributions, so that only the leading term with $r_0$ absorbed photons must be taken into account. The rate \eqref{rate} may be transformed into the laboratory system according to
\begin{equation}\label{Rtrafo}
 R_{\text{\tiny Lab}}=\frac{R}{\gamma_{\text{\tiny Lab}}}\,,
\end{equation}
where $\gamma_{\text{\tiny Lab}}$ is the Lorentz factor of the incoming muon in the laboratory system. Note that total probabilities of multiphoton processes are usually expressed directly as rates rather than cross sections, because the cross section $\sigma_r=R_r/j$ of a nonlinear process ($r>1$) still depends on the incoming photon flux $j\propto\xi^2$, which is an undesired feature for a cross section in the usual sense.




\section{Results}\label{Resultsec}
\subsection{Total pair creation rates}\label{TotalRates}

In this section, we present the results of our numerical calculations of the total pair production rates. They refer to the rates of produced electron--positron pairs per projectile muon in the restframe of the incoming muon for a laser beam of infinite length. In order to obtain the total pair yields in the laboratory system, which could be observed in a corresponding experiment, one has to transform into the laboratory frame (see Eq. \eqref{Rtrafo}) and multiply the outcoming rate by the interaction time. The latter is given by half the laser pulse duration $\tau$, provided that this is considerably small compared to the muon lifetime, which usually is the case (typically, $\tau\sim\text{fs}-\text{ns}$, while the muon lifetime in the lab frame is $\tau_\mu=\gamma_{\text{\tiny Lab}}\cdot2\mu$s). Finally, one has to multiply by the number of muons in the projectile beam. 

\subsubsection{Linear process}
For linear processes, the corresponding expressions for the total rates can be found using the framework of standard QED \cite{LandauLifschitz, MotzRev, Jauch}. This allows for a testing of our calculation by comparison.
\\We first consider an $e^+e^-$ pair creation process by a single photon of the energy $\omega=1.8$MeV impinging on a muon initially at rest. 
We choose the laser intensity parameter to be $\xi=7.5\times 10^{-4}$, which corresponds to XFEL photons with a lab frame energy of $\omega_{\text{\tiny Lab}}=9$keV and an intensity of $I_{\text{\tiny Lab}}=8\times10^{19}$W/cm$^2$, which are envisaged to become available by the X--ray lasers planned at SLAC (Stanford, USA) and DESY (Hamburg, Germany) \cite{SLAC:DESY}. In this constellation, a muon Lorentz factor of $\gamma_{\text{\tiny Lab}}=100$ would be necessary. Alternatively, if even higher muon energies were to become accessible \cite{Alsharoa, AAkesson, MCTF-Report}, the assumed set of parameters could be implemented by an XUV laser with a photon energy of $\omega_{\text{\tiny Lab}}=90$eV and an intensity of $I_{\text{\tiny Lab}}=8\times 10^{15}$W/cm$^2$ \cite{Charalambidis} with a muon Lorentz factor of $\gamma_{\text{\tiny Lab}}=10^{4}$. The numerical calculation of the total rate in the rest frame of the incoming muon \eqref{totRate} yields for these parameters  
\begin{equation}\label{onePhot_muon}
 R_{r=1}(\omega=1.8\text{MeV})= 1.28\times 10^9 \text{s}^{-1}\,.
\end{equation}
This rate results from a single term in the double sum of Eq. \eqref{IntegrationCM} with $r=n=n'=1$, i.~e. one laser photon is absorbed at the electron--positron vertex, and no photon is absorbed (or emitted) at the projectile vertex. This term comprises the first two Feynman diagrams on the right--hand side in Fig. \ref{FeynDiagExpansion1}. All other diagrams yield negligibly small contributions.
\\The rate is proportional to $\xi^2$. In mathematical terms this scaling arises from the Bessel function $J_n(\bar\alpha)$ in Eq. \eqref{sp-mat-pr}. Since $\bar\alpha\ll1$ and $\bar\alpha\propto\xi$, the amplitude for one--photon absorption at the $e^+e^-$ vertex is proportional to $\xi$, leading to $R_{r=1}\propto\xi^2$. 
The properties of the Bessel functions also explain why terms involving photon exchange at the muon vertex are strongly suppressed and may be ignored. Those terms are proportional to $J_N(\bar\beta)\propto\bar\beta^{\abs{N}}$ with $\bar\beta\propto\Xi=\xi m/M $, leading to contributions to the process rate which are smaller than the leading term ($N=0$) by at least three orders of magnitude. 
%
Higher photon orders yield even smaller contributions. 
%
\\Since the recoil energy is small compared with $Q^0\approx M$, the main contribution comes from the $\mu=\nu=0$ term in the sum over the spinor--matrix product \eqref{sp-mat-pr} (see also App. \ref{compApp}). We found this term to yield $97.5 \%$ of the given result \eqref{onePhot_muon}.\\
%
We expect the numerical value of the total pair production rate for muon projectiles to be approximately the same as for e.~g. proton projectiles because both particles are very heavy as compared to the electron mass.  
Since protons are not fundamental particles, our calculation does not actually apply to them. But since they are spin--$\nicefrac{1}{2}$ particles like leptons, we may approximately treat them as effective Dirac particles and check the result of our numerical program code for projectiles having the mass of a proton. This yields a total pair production rate of {$R_{1}^{M=m_p}(1.8$MeV$)=1.31\times10^9$s$^{-1}$}, which is in agreement with the result obtained from the calculation treating protons as external Coulomb fields \cite{Milstein06}. 
The difference between the total rate for muon and proton projectiles stems from the recoil effects due to the lower mass of the former particle.\\

If the absorbed energy is very close to the threshold energy \eqref{threshold}, i.~e. the photon energy in an $r$--photon process is close to
\begin{equation}\label{wmindef}
 \omega_{r,\text{\tiny min}}=\frac{2m}{r}(1+\frac{m}{M})\,,
\end{equation}
%
%
one finds a power law for the dependence of the total pair creation rate on the amount of energy absorbed additionally to ${\omega_{r,\text{\tiny min}}}$:
\begin{equation}\label{scaling}
 R_r\propto(\omega-\omega_{r,\text{\tiny min}})^{\chi_r}\,.
\end{equation}
For one--photon $e^+e^-$ pair production processes, our numerical results for muon projectiles can be fitted to the functional relation of Eq. \eqref{scaling} and yield the exponent
\begin{equation}
 \chi_1=2.97,
\end{equation}
with an accuracy of $0.1\%$ within the range between $\omega_{1,\text{\tiny min}}$ and $1.05\times\omega_{1,\text{\tiny min}}$. In the case of very heavy projectile particles, the exponent is $\chi_1^{M\rightarrow\infty}=3$ \cite{LandauLifschitz}. An investigation of the scaling behavior for proton projectiles with our program code yielded agreement with this value. For electron projectiles, the exponent is $\chi_1^{m_e}=2$ \cite{Jauch}. 
We do not consider electron projectiles here, since our calculation does not take into account the quantum mechanical exchange term of the produced and scattered electron. But we may calculate the rates for hypothetical projectile particles carrying e.~g. twice the electron mass. While the minimum absorbed energy $\omega_{1,\text{\tiny min}}$ is $4m_e$ for electron projectiles, and approximately $2m_e$ for protons and also for muons, for hypothetical projectile particles with $M=2m_e$, the threshold energy lies just in the middle at $3m_e$. For such particles, it is not possible to neglect summands with $N,N'\neq0$ in \eqref{IntegrationCM}. Instead, we employed the two leading orders, $n=n'=1$ and $n=n'=0$. We find the exponent to be $\chi_1^{M=2m_e}=2.02$ with an accuracy of $0.2\%$ between $\omega_{1,\text{\tiny min}}$ and $1.06\times\omega_{1,\text{\tiny min}}$. 
This exponent is close to the one obtained for electron projectiles. 


\subsubsection{Nonlinear processes}

The same amount of energy as in the one--photon process considered in the foregoing paragraph is absorbed in a two--photon process with half the photon energy, i.~e. $\omega=900$keV in the incoming muon's rest frame. Using the same laser intensity parameter as in the linear case, we obtain the total pair production rate for the two--photon process with muon projectiles,
\begin{equation}\label{twophot_mum}
 R_{r=2}(\omega=900\text{keV})= 187.5 \text{s}^{-1}\,.
\end{equation}
This is considerably less than obtained for the absorption of one single laser photon with twice the energy. 
Since the pair production rate scales with $\xi^{2r}$, we expect the rate for the two--photon process to be roughly 7 orders of magnitude smaller than for the one--photon process, which agrees with Eqs. \eqref{onePhot_muon} and \eqref{twophot_mum}. Also here, the main contribution to the pair production rate comes from the $\mu=\nu=0$ summand in the trace product \eqref{SpinSum} according to \eqref{00term}. In this case the said summand yields $97.8\%$ of the total rate \eqref{twophot_mum}. The calculation neglecting the finite proton mass yields for this parameter constellation the value of {$R_{2}^{M\rightarrow\infty}(\omega=900$keV$)= 190.0 \text{s}^{-1}$} \cite{Milstein06}.\\
%
%

\begin{figure}
 \includegraphics[width=0.4\textwidth]{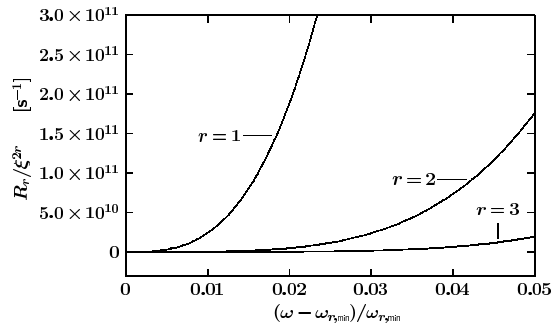}\caption{Pair creation rates $R_r$ for muon projectiles as a function of the energy absorbed additionally to $\omega_{r,\text{\tiny min}}$ of Eq. \eqref{wmindef} for photon orders 
$r=1,2,3$. The exponent increases as the number of absorbed photons increases.}\label{scaleplot}
\end{figure}

As in the foregoing paragraph, it is possible to find a scaling behavior of the total pair production rate for total absorbed energies close to the threshold energy, see Eq. \eqref{scaling}. For two--photon processes we find the exponent
\begin{equation}
\chi_2=3.92
\end{equation}
for muon projectiles with a relative error of $0.1\%$ between $\omega_{2,\text{\tiny min}}$ and $1.05\times\omega_{2,\text{\tiny min}}$. Considering proton projectiles, we find the exponent to be $3.96$ with an error of $0.23\%$, which agrees with the result in \cite{Milstein06}, and for hypothetical particles with twice the electron mass, the exponent is $3.02$ with the accuracy of $0.3\%$. \\We observe that in each case, the exponents exceed the ones found in the foregoing paragraph by about $1$. An investigation of higher photon orders up to $r=5$ shows that the increase of the exponent is proportional to the increase of the photon order (see Fig. \ref{scaleplot}). We find
\begin{equation}
\chi_r\approx1.95+r\,
\end{equation}
for muon projectiles.
\\It should be noted that regarding this scaling behavior, the circular polarization of the laser is crucial; for linear polarization, the scaling would be different in nonlinear processes. There one finds $\chi_2^{M\rightarrow\infty}=2$, for instance \cite{Milstein06}. Polarization dependence is a characteristic feature of nonlinear processes, in general.
\\

As in the one--photon process considered above, only $N=N'=0$ summands significantly contribute to the total pair creation rate, i.~e. the absorption or emission of laser photons at the muon vertex are suppressed. We note however, that so--called resonances can occur in the terms involving photon exchange both at the electron--positron vertex and at the muon vertex (${n\neq0,N\neq0}$). Then, for certain final--state momenta, the virtual photon may reach the mass shell, ${q^2=(q_++q_- -nk)^2=0}$, and become a real photon. Physically, this resonant behavior means that the original second--order Feynman graph in Fig. \ref{FeynDiag1} decomposes into two first--order processes: first the emission of a real photon by a muon and second the production of an $e^+e^-$--pair by a real (non--laser) photon (Breit--Wheeler process). Both processes are kinematically allowed in the presence of an external laser field; the second process was responsible for the pair production observed in the SLAC experiment \cite{SLAC,SLAC2}. In our case however, the real--photon channel is strongly suppressed by the large muon mass: while the process involving the propagation of a virtual photon with no absorption at the muon vertex yields a production rate proportional to $\xi^4$ for a two--photon process, the corresponding rate for pair production involving the absorption of one photon at the muon vertex is proportional to $\xi^2\Xi^2\ll\xi^4$. Resonances in multiphoton processes have mainly been discussed for the examples of laser--assisted $e^-e^-$ (M\o ller) scattering \cite{Panek}, $e^+e^-$ (Bhabha) scattering \cite{Denisenko}, and electron bremsstrahlung \cite{ErikPropagator}.

\subsection{Recoil effects}

\begin{figure}
 \includegraphics[width=0.4\textwidth]{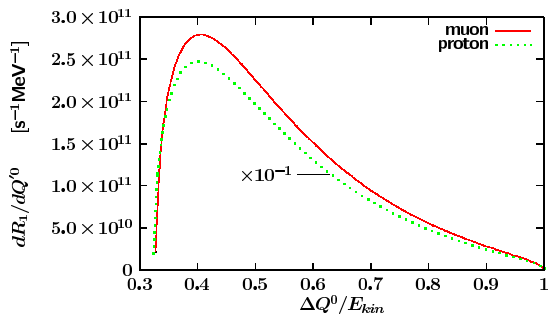}\caption{(Color online.) Differential pair creation rate plotted against the energy loss of the projectile relative to its initial kinetic energy for the one--photon process at $\omega=1.8$MeV. The red solid line shows the recoil distribution for muons, the green dotted curve shows the one for proton projectiles. For protons, the differential rate has been divided by a factor 10.}\label{rec1pnonscaled}
\end{figure}

In this section, we discuss the recoil distributions in the c.m. system for the already considered one- and two--photon processes, and also for a three--photon process, for different projectile particles. 
\\Figure \ref{rec1pnonscaled} shows the c.m. differential rate for muon and proton projectiles plotted against the energy loss relative to the incoming projectile's kinetic energy. Note that for the protons, we divided the differential rate by a factor 10 in order to fit into the graph. 
In the c.m. system, the incoming muon projectile has a kinetic energy of ${E_{kin}=14.83}$keV, and after the collision with the laser beam, the most probable kinetic energy is ${E'_{kin}=8.81}$keV, i.~e. the muon loses about $40.6\%$ of its initial energy of motion. The proton projectile possesses an initial kinetic energy of ${E_{kin}=1.72}$keV in the c.m. system, and $E'_{kin}=1.03$keV after the collision; i.~e. it loses about $40.2\%$ of its initial energy of motion. The absolute loss of energy is about ${E_{kin}-E'_{kin}=:\Delta Q^0=6}$keV for muons and ${\Delta Q^0 = 0.7}$keV for protons, which is in agreement with the general rule that the recoil momentum $\Delta \abs{\bm Q}$ should be of the order of $2m$. Since the heavy projectiles move nonrelativistically, this should lead to a recoil energy $\Delta Q^0\sim\Delta \abs{\bm Q}^2/2M$, and therefore the absolute energy loss for the muon should be one order of magnitude larger than for the proton. As discussed in Section \ref{TotalRates} and in Appendix \ref{compApp}, the total pair creation rates for muon and proton projectiles yield almost the same value. As shown in Fig. \ref{rec1pnonscaled}, the shapes of the recoil distribution for the two projectile species are similar. The distribution for the proton is about a factor $10$ higher than the one for muons, but it is also a factor 10 narrower if absolute energy loss is considered, so that the integration leads to about the same numerical value for both projectile types.
%
\begin{figure}
\centering
\includegraphics[width=0.45\textwidth]{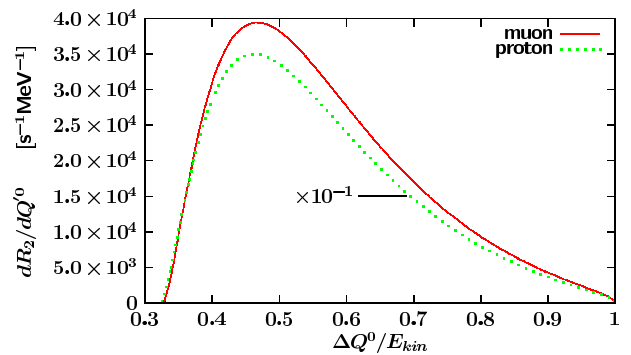}\caption{(Color online.) Same as Fig. \ref{rec1pnonscaled}, but for a two--photon process with $\omega=900$keV.
}\label{rec2pnonscaled}
\end{figure}
\\If the same amount of energy is absorbed in a two--photon process, as considered above, the shape of the recoil distribution is quite similar to the one for the one--photon process in Fig. \ref{rec1pnonscaled}, but slightly broader and shifted to higher relative energy loss (Fig. \ref{rec2pnonscaled}). The initial kinetic energy of the projectile particles in the c.m. system is the same as in the one--photon process, since the total absorbed energy is the same in both cases. Here, for the muon the most probable kinetic energy after the impact of the laser beam is $7.93$keV which is tantamount to a loss of $46.5\%$ of the initial kinetic energy. The proton projectile loses $46.1\%$ of its initial kinetic energy and ends up with a kinetic energy of $0.93$keV.\\
In both the linear and the nonlinear process, both projectile species lose about the same ratio of their initial energy of motion in the collision with the laser photons. But although the total amount of absorbed energy, as well as the initial energy of the projectile in the c.m. system and the upper limit \eqref{Qprime_max} for its final energy are the same in both considered processes, the lost energy is significantly larger in the nonlinear, two--photon process. 
\\This trend is confirmed by the investigation of another kinematically identical process yielding the same value of the totally absorbed energy from the laser field: if three photons of energy $600$keV are absorbed, the scattered muon's most probable kinetic energy is $7.49$keV, which corresponds to a loss of $49.5\%$ of the initial kinetic energy. The most probable kinetic energy of a scattered proton projectile lies at $0.88$keV, which means a loss of $49.1\%$ of the initial kinetic energy. It should be noted that the considered three--photon process is unlikely to be experimentally observed, because for the considered photon energy, also a two--photon process is possible. Because $R_r\propto \xi^{2r}$, the two--photon process is $7$ orders of magnitude more probable than the three--photon process.\\
We point out that the increasing of the energy loss of the projectile is in agreement with the fact that the emission angles of the produced pair rise with the photon order \cite{Carsten}. The angular distribution of the electron and positron is mainly governed by the Bessel function $J_n(\bar\alpha)$, where $\bar\alpha\propto\sin\vartheta_{q_\pm}$ (see Eq. \eqref{alphadef}). Due to the small--argument behavior of $J_n(\bar\alpha)$, large values of $n$ favor large emission angles $\vartheta_{q_\pm}$. Since the typical energy of the electron and positron is not much affected by the photon order (see Fig. \ref{qmdis} below), this implies that the longitudinal component of the scattered muon momentum decreases when $n$ (i.~e. $r$) increases (see also Fig. \ref{kinvis} below).
\\

\begin{figure}
 \includegraphics[width=0.4\textwidth]{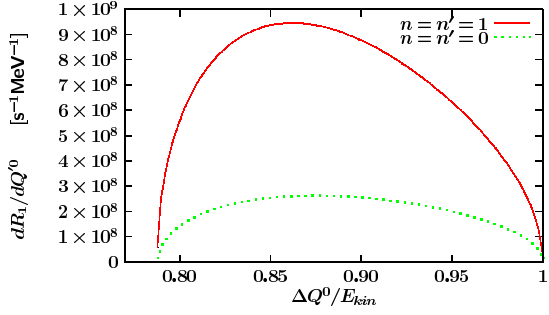}\caption{(Color online.) Recoil distribution for hypothetical projectile particle with mass $M=2m$ for a one--photon process at $\omega=1.8$MeV. The two most important orders, $n=n'=1$ (red solid line) and $n=n'=0$ (green dotted line) are depicted.}\label{rec2phypmass}
\end{figure}

In order to demonstrate that our numerical calculation yields larger recoil effects for lighter particles, we consider once more the case of a hypothetical particle possessing twice the electron mass. Figure \ref{rec2phypmass} shows the recoil distribution for the main contributions to the total rate stemming from $n=n'=1$ and $n=n'=0$ for the case of a one--photon process at the photon energy $\omega=1.8$MeV in the restframe of the incoming projectile. Initially, the hypothetical particle possesses a kinetic energy of $305$keV. The total recoil distribution consists of the sum of the two depicted orders and has its maximum at $263$keV, i.~e. the particle loses $86.4\%$ of its initial kinetic energy. As expected, the lighter particle loses a higher ratio of its initial energy in the collision with the laser beam. Also here, our calculation of the corresponding two--photon process shows that the energy loss is larger in the nonlinear than in the linear process.

%
%


\subsection{Differential rates}

In this paragraph, we discuss some further differential pair production rates for muon projectiles. Each of the following figures is to be understood in the c.m. system for the respective $r$--photon process. As in the foregoing paragraph, we consider a one--photon process at $\omega=1.8$MeV, a two--photon process at $\omega=900$keV, and a three--photon process at $\omega=600$keV. 
\\
\begin{figure}
\includegraphics[width=0.4\textwidth]{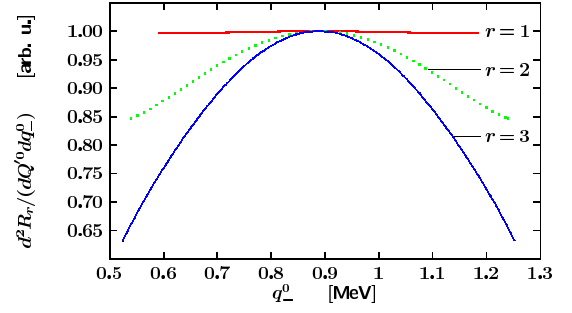}\caption{(Color online.) (Normalized) double differential pair creation rate in arbitrary units as function of the energy of the produced electron for muon projectiles. The thick red line shows the distribution for the one--photon process at $1.8$MeV, the green dotted line shows it for the two--photon process at $900$keV, and the thin blue line for the three--photon process at $600$keV. The kinetic energy of the scattered muon is set to the most probable value, i.~e. $8.8$keV for the one--photon, $7.9$keV for the two--photon, and $7.5$keV for the three--photon process. The distributions have been divided by their maximum values in order to show the evolution of the shape with increasing photon order.}\label{qmdis}
\end{figure}

Fig. \ref{qmdis} shows the dependency of the (normalized) differential pair creation rate on the energy of the produced electron. The most probable electron energy
, at which the maximum occurs, is about $890$keV for all considered photon orders. It slightly decreases with increasing photon order, but the difference between the most probable electron energy for the one- and the three--photon process is only $0.7$keV.\\
The differential rates in Fig. \ref{qmdis} have been divided by their respective maximum values. This allows for a comparison of the distributions for the different photon orders. The curvature of the distribution increases with the number of absorbed laser photons: for the considered one--photon process at $\omega=1.8$MeV, the relative height of the maximum is only $0.4\%$ of its absolute value. Considering the two--photon process at $\omega=900$keV, we find a similar shape, but here the relative height of the maximum is $15\%$ of its value, and for the three--photon process, the relative height is $37\%$. The reason is that in processes of high photon order, the $e^+e^-$ pair is preferentially produced in a symmetric configuration with ${q_+^0\approx q_-^0\approx(E-Q'^0)/2}$ and $\vartheta_{+}\approx\vartheta_-$ (see Ref. \cite{Muller03a}). Therefore, deviations from the maximum in the middle of the allowed energy range yield smaller contributions and the maximum becomes more pronounced.\\
\begin{figure}
\includegraphics[width=0.4\textwidth]{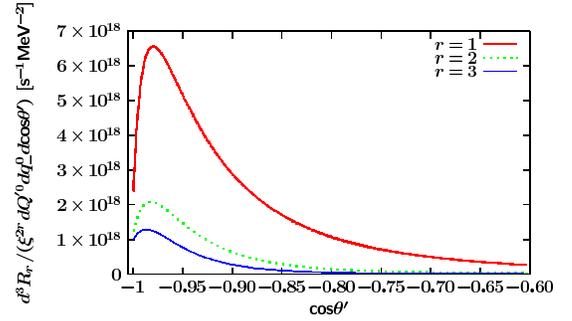}\caption{(Color online.) Dependency of the differential pair creation rate on the scattered muon's polar angle. The respective photon energies and the corresponding scattered muon's energies are as in Fig. \ref{qmdis}. The energy of the produced electron has been set to the most probable values, which is $q_-^0\approx890$keV in all cases.}\label{cosdis}
\end{figure}

\begin{figure}
 \includegraphics[width=0.3\textwidth]{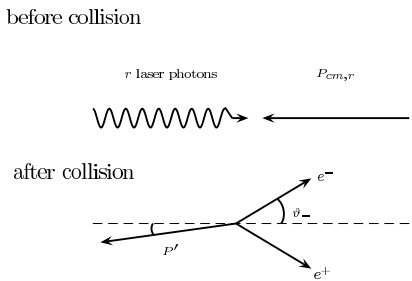}\caption{Visualization of the kinematics in the c.m. system. The $r$ laser photons are depicted as one ''big'' photon. The vanishing total momentum in this frame leads to a corresponding correlation of the particle momenta in the final state.}\label{kinvis}
\end{figure}

Fig. \ref{cosdis} shows the dependency of the differential pair production rate on the cosine of the scattered muon's polar angle for the considered one-, two-, and three--photon processes. The higher the number of absorbed laser photons, the less the muon is deflected from its initial direction of motion along the negative $z$--direction. (Note that the laser beam propagates in positive $z$--direction, see Fig. \ref{kinvis}.) This behavior is best understood by considering the argument used to explain the increasing curvature of the produced electron's energy distribution with increasing photon order. Since the symmetry between the produced electron and positron emission angles becomes more pronounced for higher photon orders, the transverse momentum of the scattered muon and thus its deflection angle are reduced (see Fig. \ref{kinvis}). \\Fig. \ref{kinvis} can also help us visualize the aforementioned increase of the muon's energy loss. It has been said that for large photon orders $n=r$, the Bessel functions $J_n(\bar\alpha)$ favor the emission of the $e^+e^-$ pair under large emission angles. The typical energies of the electron and positron are unaffected by the photon order (Fig. \ref{qmdis}). Since the emission angles of the electron and positron get larger while their energies stay the same, the longitudinal component of the sum of the pair particles' momenta gets smaller for higher photon orders. Therefore, as can be seen in Fig. \ref{kinvis}, the scattered muon's longitudinal momentum projection must decrease with increasing photon order.

\section{Conclusions}\label{ConclusionSec}

In this paper we considered the creation of electron--positron pairs in the head--on collision of a relativistic muon and a high--frequency laser beam in the multiphoton regime. 
\\The results for the total pair creation rates for linear and nonlinear pair production have been compared to the results known for projectiles of infinite mass. Close to the threshold energy, we derived a power law for the pair creation rates' dependence on the absorbed energy surmounting the threshold. We found the exponent to increase linearly with the number of absorbed photons. 
The corresponding recoil distributions have been compared to the ones of protons as effective Dirac particles and to a hypothetical lighter particle of twice the electron mass. We have shown that in the c.m. system, the projectile's loss of kinetic energy in the inelastic collision with the laser beam relative to its initial kinetic energy is approximately the same for muon and proton projectiles. For both species, we found characteristic multiphoton signatures: the projectile energy loss increases with the number $r$ of totally absorbed laser photons (with $r\omega=const$), while the deflection angle of the scattered projectile decreases.\\

Our results could be tested experimentally by combining a coherent high--energy photon source with a beam of relativistic muons. Today, muon beams with Lorentz factors of about $\gamma_{\text{\tiny Lab}}=2.8$ are produced e.~g. at PSI in Switzerland or TRIUMF in Canada \cite{PSI:TRIUMF}. Usually, they are then slowed down in order to produce e.~g. muonic atoms. In the present paper, we consider the low--intensity high--photon--energy regime of the pair production process. Therefore, we need high photon energies in the rest frame of the incoming projectile muon. Assuming a lab--frame photon energy of $9$keV, as is planned to be reached at the upcoming XFEL facilities \cite{SLAC:DESY}, a muon Lorentz factor of $\gamma_{\text{\tiny Lab}}=50$ would be needed to obtain a rest frame photon energy of $900$keV, the photon energy we assumed for the considered two--photon process. An according muon acceleration is necessary for experimental observation of the here discussed nonlinear pair production processes. Corresponding efforts are being undertaken to develop muon accelerators for e.~g. neutrino factories or muon colliders reaching muon energies up to the TeV range ($\gamma_{\text{\tiny Lab}}\sim10^4$) \cite{Alsharoa,AAkesson,MCTF-Report}.



\begin{appendix}\section{Comparison between muon and proton projectiles in the multiphoton regime}\label{compApp}

In this appendix we want to compare the present treatment of the nonlinear Bethe--Heitler process for muon projectiles to the external Coulomb field approach \cite{Yakovlev,Mittleman87,Roshchupkin9601,Dietz98,Dremin02,Avetissian03,Muller03a,Carsten,Kaminski06,Milstein06,Kuchiev07,CarlusCarsten} for proton projectiles. To this end, we first summarize the main steps and results of the latter approach.\\
The heavy projectile is taken into account only as the Coulomb field $A_C$ of a nucleus, so that a proton in its rest frame can be described by the four potential
\begin{equation}
 A^\mu_C=(\frac{e}{\abs{\bm x}},0,0,0)\,.
\end{equation}
The approximation of an infinitely heavy projectile leads to an amplitude which describes the transition of an electron from a negative- to a positive--energy state as induced by the Coulomb field of the proton:
\begin{equation}
 \mathscr{S}_{p_+p_-}=ie\int \bar\psi_{p_-,s_-} \dirac A_C \psi_{p_+,s_+}d^4x\,,
\end{equation}
with the Volkov states of Eq. \eqref{VolkovElectron}. Similarly to the expansion for the electronic vertex in the calculation of Sec. \ref{TheorySec}, this expression is expanded in a Fourier series with coefficients depending on the regular Bessel functions $J_r(\bar\alpha)$. One finds for the partial rate corresponding to the photon order $r$
\begin{equation}
 R^{(r)} = \int\frac{Vd^3q_+}{(2\pi)^3}\frac{Vd^3q_-}{(2\pi)^3} \frac{1}{T}\sum_{s_+s_-}\abs{\mathscr S^{(r)}_{p_+p_-}}^2\,,
\end{equation}
with the square of the partial transition amplitude
\begin{equation}\label{PartAmSqProton}
 \abs{\mathscr S^{(r)}_{p_+p_-}}^2=\frac{4\pi m^2e^2}{V^2q_+^0q_-^0}\abs{\mathcal{M}_{p_+p_-}^{(r)}}^2\frac{2\pi \delta(q_r^0)}{\bm{q}_r^4}\cdot T\,,
\end{equation}
where the four vector $q_r=q_++q_--rk$ represents the momentum transfer on the proton. Since the proton is assumed to be infinitely heavy, there is no energy transfer, $q_r^0=0$
. The spinor--matrix product $\mathcal{M}_{p_+p_-}^{(r)}$ here reads
\begin{align}
\mathcal{M}_{p_+p_-}^{(r)}&:=\bar{u}_{p_-s_-}\Gamma_r^\infty v_{p_+ s_+}\,,
\end{align}
with
\begin{align}
 \Gamma_r^\infty=\Bigl[&\Bigl(\gamma^0 - \frac{e^2a^2k^0\dirac k}{2(kp_+)(kp_-)} \Bigr)B_{r} \nonumber\\
&+ \frac{e}{2}\Bigl(\frac{1}{kp_+}\gamma^0\dirac k \dirac a_1 - \frac{1}{kp_-}\dirac a_1\dirac k\gamma^0\Bigr)C_{r} \nonumber\\
&+ \frac{e}{2}\Bigl(\frac{1}{kp_+}\gamma^0\dirac k \dirac a_2 - \frac{1}{kp_-}\dirac a_2\dirac k\gamma^0\Bigr)D_{r}\Bigr]\,.
\end{align}
The coefficients $B_r,$ $C_r$ and $D_r$ are defined as in Eq. \eqref{coefficients}. Note that in this approach, all the photons are absorbed at the vertex of the produced pair (since the projectile vertex is not taken into account), $r=n$, independently of the considered laser intensity regime. Because the proton is treated as an external Coulomb field, only the $\mu=0$ components of Eq. \eqref{sp-mat-pr} contribute to the transition. Thus, one finds the expression for the spin sum over the square of the spinor--matrix product
\begin{equation}\label{protonSpur}
 \sum_{s_+s_-}\abs{\mathcal M_{p_+p_-}^{(r)}}^2=\text{Tr}\bigl(\Gamma_r^\infty\frac{\dirac p_+-m}{2m}\bar\Gamma_r^\infty\frac{\dirac p_-+m}{2m} \bigr)\,.
\end{equation}
With this, we have listed the most important relations of the calculation for proton projectiles. 
%
We now show that the exact treatment for muon projectiles performed in Sec. \ref{TheorySec} is structurally similar to it in the limit of small laser intensities. For small $\Xi\ll\xi\ll1$, the effective mass and momentum of the muon read
\begin{equation}\label{Meffapprox}
 M_*=M\sqrt{1+\Xi^2}\approx M\,,\quad Q^\nu = P^\nu+\Xi^2\frac{M^2}{2kP}k^\nu \approx P^\nu\,.
\end{equation}
Because of the large muon mass, we may assume nonrelativistic momentum changes for the scattered projectile. Then, in the rest frame of the incoming muon, the energy of the scattered muon is close to its rest mass:
\begin{equation}\label{Papprox}
P'^0\approx M+\frac{|\bm{P}'|^2}{2M}\,,
\end{equation}
with $\abs{\bm{P}'}\ll M$. We now consider the argument of the Bessel functions governing the muon vertex, $\bar\beta$ from Eq. \eqref{betadef}. It can be written as
\begin{equation}
 \bar\beta=\frac{ea|\bm{P}'_\perp|}{\omega({P'_0}-{P'_z})}\approx {\xi} {\frac{m}{\omega}}{\frac{|\bm{P}'_\perp|}{M}} \ll 1\,.
\end{equation}
Since, for small arguments, the regular Bessel functions scale with $J_N(\bar\beta)\propto\bar\beta^{\abs{N}}$ \cite{Abramowitz}, all orders $N\neq0$ are negligible for very small $\bar\beta$. Thus, the double sum in \eqref{IntegrationCM} collapses to a sum over ${n=n'=r}$. This means that in the limit of nonrelativistic motion of the scattered muon within the rest frame of the incoming one, no emission or absorption of laser photons takes place at the projectile vertex. In this case, the summands in the matrix elements \eqref{DeltaDef} estimate to be negligible except for $\gamma_\mu F_0$. The Bessel function $J_0$ is about unity for small arguments, $J_0(0)=1$ \cite{Abramowitz}. Then, the muon part of the spin sum over the spinor--matrix product in \eqref{SpinSum} becomes
\begin{align}
 \sum_{S,S'}\mathcal{M}_\mu&(\mu,\mu'|0) \mathcal{M}^\dagger_\nu(\mu,\mu'|0)\nonumber\\
&
\approx \text{Tr}\Bigl(\gamma_\mu  \frac{\dirac P_-+M}{2M} {\gamma_\nu} \frac{\dirac P'_-+M}{2M}\Bigr)=2\delta_{\mu0}\delta_{\nu0}\,,
\end{align}
so that the total sum over final spins and average over initial spins \eqref{SpinSum} estimates
\begin{widetext}
\begin{align}\label{00term}
 \frac{1}{2} &\sum_{S}\sum_{S',s_+,s_-}\mathcal{M}^\mu(e^+e^-|n)\mathcal{M}_\mu(\mu,\mu'|r-n) \mathcal{M}^\dagger_\nu(\mu,\mu'|r-n')\mathcal{M}^{\dagger\nu}(e^+e^-|n')\nonumber\\
&=\frac{1}{2}\sum_{S}\sum_{S',s_+,s_-}\mathcal{M}^0(e^+e^-|r)\mathcal{M}_0(\mu,\mu'|0) \mathcal{M}^\dagger_0(\mu,\mu'|0)\mathcal{M}^{\dagger0}(e^+e^-|r)\approx \text{Tr}\left(\Gamma_r^0\frac{\dirac p_+ - m}{2m}\bar{\Gamma}_{r}^0 \frac{\dirac p_- + m}{2m}\right),
\end{align}
\end{widetext}
which is just the expression obtained for projectiles of infinite mass, Eq. \eqref{protonSpur}.
\\Another formal difference between the two approaches is that in the present calculation, we divide the transition amplitude by the four--momentum of the virtual photon \eqref{partAmpli}, whereas in the external Coulomb field approach the division by the virtual photon's three--momentum is performed \eqref{PartAmSqProton}. %
%
%
Performing the space--time integral over the projectile coordinate in \eqref{Ampli} under the assumption of no photon emission or absorption at the projectile vertex, the obtained four--dimensional $\delta$--function (see Eq. \eqref{transAmpl}) reads $\delta(Q+q-Q')$. With the approximation \eqref{Papprox}, only $\delta(q^0)$ remains in the $0$--component of this $\delta$--function (cf. Eq. \ref{PartAmSqProton}). Therefore, ${\abs{q^2}}=\bm{q}^2$, and the difference between the two approaches concerning the division by the square of the virtual photon's momentum vanishes in the considered limit (see also \cite{PeskinSchroeder}, Sec. 4.8 for a similar argument). In the present calculation, the remaining three--dimensional $\delta$--function allows for the additional integration over the scattered projectile's momentum to be carried out.
\end{appendix}


\begin{thebibliography}{99}

\bibitem{Sauter}
F. Sauter, Z. Phys. \textbf{69}, 742 (1931).
\bibitem{Marklund}
M. Marklund, P. Shukla, Rev. Mod. Phys. \textbf{78}, 591 (2006).
\bibitem{Salamin06}
Y.~I. Salamin, S.~X. Hu, K.~Z. Hatsagortsyan, and C.~H. Keitel, Phys. Rep. \textbf{427}, 41 (2006).
\bibitem{EhlotzkyRev}
F. Ehlotzky, K. Krajewska, and J.~Z. Kami\'nski, Rep. Prog. Phys. \textbf{72}, 046401 (2009).
\bibitem{Schwinger}
J. Schwinger, Phys. Rev. \textbf{82}, 664 (1951).

\bibitem{Brezin70}
E. Brezin and C. Itzykson, Phys. Rev. D \textbf{2}, 1191 (1970).
\bibitem{Popov71}
V.~S. Popov, JETP Lett. \textbf{13}, 185 (1971).
\bibitem{Schuetzhold08}
R. Sch\"utzhold, H. Gies, and G. Dunne, Phys. Rev. Lett. \textbf{101}, 130404 (2008).
\bibitem{Rufus}
M. Ruf, G.~R. Mocken, C. M\"uller, K.~Z. Hatsagortsyan, and C.~H. Keitel, Phys. Rev. Lett. \textbf{102}, 080402 (2009).
\bibitem{Cheng}
T. Cheng, Q. Su, and R. Grobe, EPL \textbf{86}, 13001 (2009).

\bibitem{Reiss62}
H.~R. Reiss, J. Math. Phys. \textbf{3}, 59 (1962).
\bibitem{Reiss71}
H.~R. Reiss, Phys. Rev. Lett. \textbf{26}, 1072 (1971).
\bibitem{Narozhnyi}
N.~B. Narozhnyi, A.~I. Nikishov, and V.~I. Ritus, Zh. Eksp. Teor. Fiz. \textbf{46}, 776 (1964) [Sov. Phys. JETP \textbf{19}, 529 (1964)].
\bibitem{Bell}
A.~R. Bell, J.~G. Kirk, Phys. Rev. Lett. \textbf{101}, 200403 (2008).


\bibitem{BetheHeitler}
H. Bethe and W. Heitler, Proc. Roy. Soc. London A \textbf{146}, 83 (1934).
\bibitem{BreitWheeler}
G. Breit and J.~A. Wheeler, Phys. Rev. \textbf{46}, 1087 (1934).

\bibitem{SLAC}
D.~L. Burke, R.~C. Field, G. Horton-Smith, J.~E. Spencer, D. Walz, S.~C. Berridge, W.~M. Bugg, K. Shmakov, A.~W. Weidemann, C. Bula, K.~T. McDonald, E.~J. Prebys, C. Bamber, S.~J. Boege, T. Koffas, T. Kotseroglou, A.~C. Melissinos, D.~D. Meyerhofer, D.~A. Reis, and W. Ragg, Phys. Rev. Lett. \textbf{79}, 1626 (1997).
\bibitem{SLAC2}
C. Bamber, S. J. Boege, T. Koffas, T. Kotseroglou, A. C. Melissinos, D. D. Meyerhofer, D. A. Reis, W. Ragg, C. Bula, K. T. McDonald, E. J. Prebys, D. L. Burke, R. C. Field, G. Horton-Smith, J. E. Spencer, D. Walz, S. C. Berridge, W. M. Bugg, K. Shmakov, and A. W. Weidemann, Phys. Rev. D \textbf{60}, 092004 (1999).
\bibitem{Yakovlev}
V.~P. Yakovlev, Zh. Eksp. Teor. Fiz. \textbf{49}, 318 (1965) [Sov. Phys. JETP \textbf{22}, 223 (1966)].


\bibitem{Mittleman87} M.~H. Mittleman, Phys. Rev. A \textbf{35}, 4624 (1987).

\bibitem{Roshchupkin9601} S.~P. Roshchupkin, Laser Phys. \textbf{6}, 837 (1996);
\\Phys. At. Nucl. \textbf{64}, {243} ({2001}).

\bibitem{Dietz98} K.~Dietz and M.~Pr{\"o}bsting, J. Phys. B \textbf{31}, L409  ({1998}).

\bibitem{Dremin02} I.~M. Dremin, Pis'ma Zh. Eksp. Teor. Fiz. \textbf{76}, 185 (2002) [JETP Lett. \textbf{76}, 151 (2002)].

\bibitem{Avetissian03} {H.~K.} {Avetissian}, {A.~K.} {Avetissian}, {G.~F.} {Mkrtchian},
and {Kh.~V.} {Sedrakian}, Nucl. Instrum. Meth. Phys. Res. A \textbf{507}, {582} ({2003}).

\bibitem{Muller03a} {C.}~{M\"{u}ller}, {A.~B.} {Voitkiv}, and {N.}~{Gr\"un}, {Phys. Rev. A} \textbf{{67}}, 063407 ({2003});
\\Nucl. Instrum. Meth. Phys. Res. B \textbf{205}, 306 (2003).


\bibitem{Carsten} {C.}~{M\"{u}ller}, {A.~B.} {Voitkiv}, and {N.}~{Gr\"{u}n}, Phys. Rev. A \textbf{70}, 023412 (2004);
\\Phys. Rev. Lett. \textbf{91}, 223601 (2003).

\bibitem{Kaminski06} {J.~Z.} {Kami\'nski}, {K.}~{Krajewska}, {and} {F.}~{Ehlotzky}, Phys. Rev. A \textbf{74}, 033402 (2006);
\\{K.}~{Krajewska}, {J.~Z.} {Kami\'nski}, and {F.}~{Ehlotzky}, Laser Phys. \textbf{16}, 272 (2006);
\\{P.}~{Sieczka}, {K.}~{Krajewska}, {J.~Z.} Kami\'nski, {P.}~{Panek}, {and} {F.}~{Ehlotzky}, Phys. Rev. A \textbf{73}, 053409 (2006);
\\{K.}~{Krajewska} {and} {J.~Z.} {Kami\'nski}, Laser Phys. \textbf{18}, 185 (2008).

\bibitem{Milstein06} {A.~I.} {Milstein}, {C.}~{M\"{u}ller}, {K.~Z.} {Hatsagortsyan}, {U.~D.} {Jentschura}, {and} {C.~H.} {Keitel}, Phys. Rev. A \textbf{73}, 062106 (2006).

\bibitem{Kuchiev07} {M.~Y.} {Kuchiev} {and} {D.~J.} {Robinson}, Phys. Rev. A \textbf{76}, 012107 (2007).

\bibitem{CarlusCarsten} C. M\"uller, C. Deneke, and C.~H. Keitel, Phys. Rev. Lett. \textbf{101}, 060402 (2008);
\\C. M\"uller, Phys. Lett. B \textbf{672}, 56 (2009).

\bibitem{Tinsley05}
{T.~M.}~{Tinsley}, Phys. Rev. D \textbf{71}, 073010 (2005).

\bibitem{ldQED}
V.~I. Ritus, J. Sov. Laser Res. \textbf{6}, 497 (1985);
\\A.~I. Nikishov, J. Sov. Laser Res. \textbf{6}, 619 (1985).

\bibitem{Nedoreshta}
V~ N. Nedoreshta and A.~I. Voroshilo, Las. Phys. Lett. \textbf{4}, 12 (2007).
\bibitem{CarstenMuonpairs}
C. M\"uller, K.~Z. Hatsagortsyan, and C.~H. Keitel, Phys. Rev. D \textbf{74}, 074017 (2006);
\\Phys. Lett. B \textbf{659}, 209 (2008);
\\Phys. Rev. A \textbf{78}, 033408 (2008).
\bibitem{Thoma09}
M.~H. Thoma, Rev. Mod. Phys. \textbf{81}, 959 (2009).
\bibitem{KuchievMuonpairs}
M.~Y. Kuchiev, Phys. Rev. Lett. \textbf{99}, 130404 (2007).
\bibitem{NedoPairs}
V.~N. Nedoreshta, S.~P. Roshchupkin, and A.~I. Voroshilo, Las. Phys. \textbf{19}, 531 (2009).
\bibitem{Panek}
P. Panek, J.~Z. Kami\'nski, and F. Ehlotzky, Phys. Rev. A \textbf{69}, 013404 (2004) and references therein.
\bibitem{Denisenko}
O.~I. Denisenko and S.~P. Roshchupkin, Laser Phys. \textbf{9},
1108 (1999).
\bibitem{Erik}
E. L\"otstedt, U.~D. Jentschura, and C.~H. Keitel, Phys. Rev. Lett. \textbf{101}, 203001 (2008).

\bibitem{Toni1}
{A.} Di~Piazza, K.~Z. Hatsagortsyan, and C.~H. Keitel, Phys. Rev. Lett. \textbf{100}, 010403 (2008).
\bibitem{Toni2}
{A.} Di~Piazza and A.~I. Milstein, Phys. Rev. A \textbf{77}, 042102 (2008).

\bibitem{Beiersdorfer}
H. Chen, S.~C. Wilks, J.~D. Bonlie, E.~P. Liang, J. Myatt, D.~F. Price, D.~D. Meyerhofer, and P. Beiersdorfer, Phys. Rev. Lett. \textbf{102}, 105001 (2009).



\bibitem{Volkov}
D.~M. Volkov, Z. Phys. \textbf{94}, 250 (1935).
\bibitem{LandauLifschitz}
V.~B. Berestetskii, E.~M. Lifshitz, and L.~P. Pitaevskii,
\textit{Relativistic Quantum Theory} (Pergamon, New York,
1971).

 \bibitem{BjorkenDrell}
J.~D. Bjorken and S.~D. Drell, \textit{Relativistic Quantum Mechanics}
(McGraw-Hill, New York, 1964).
\bibitem{Becker75}
W. Becker and H. Mitter, J. Phys. A \textbf{8}, 1638 (1975); V.~N.
Baier et~al., Zh. Eksp. Teor. Fiz. \textbf{69}, 783 (1975) [Sov. Phys.
JETP \textbf{42}, 400 (1976)].
\bibitem{ErikPropagatorFootn}
Field--induced modifications of the electron propagator appear
at lower intensities already. For the case of laser--assisted
bremsstrahlung, they are discussed in \cite{ErikPropagator}.
\bibitem{ErikPropagator}
E. L{\"o}tstedt, U.~D. Jentschura, and C.~H. Keitel, Phys. Rev. Lett. \textbf{98}, 043002 (2007).

\bibitem{Abramowitz}
M. Abramowitz and I.~A. Stegun, \textit{Handbook of
Mathematical Functions} (Dover, New York, 1965).
\bibitem{Mork}
K.~J. Mork, Physica Norvegica \textbf{5}, 51 (1971).



\bibitem{MotzRev}
J.~W. Motz, H.~A. Olsen, and H.~W. Koch, Rev. Mod. Phys. \textbf{41}, 581 (1969).

\bibitem{Jauch}
J.~M. Jauch, and F. Rohrlich, \textit{The Theory of Photons and Electrons 2nd ed.} (Springer, New York, 1976).
\bibitem{SLAC:DESY}
L.~F. DiMauro \textit{et~al.}
, J. Phys. Conf. Ser. \textbf{88}, 012058 (2007);\\
M. Altarelli \textit{et~al.}, Technical Design Report of the European XFEL, DESY 2006-097, http://xfel.desy.de/tdr/tdr/.


\bibitem{Alsharoa}
M.~M. Alsharo'a \textit{et~al.}, Phys. Rev. ST Accel. Beams \textbf{6}, 081001 (2003).
\bibitem{AAkesson}
T. \AA kesson \textit{et~al.}, Eur. Phys. J. C \textbf{51}, 421 (2007).
\bibitem{MCTF-Report}
S. Greer, AIP Conf. Proc. \textbf{981}, 287 (2008).
\bibitem{Charalambidis}
D. Charalambidis, P. Tzallas, E.~P. Benis, E. Skantzakis, G. Maravelias, L.~A.~A. Nikolopoulos, A. Peralta Conde, and G.~D. Tsakiris, New J. Phys. \textbf{10}, 025018 (2008).
\bibitem{PSI:TRIUMF}
For current information see http://www.psi.ch and http://www.triumf.ca, respectively.

\bibitem{PeskinSchroeder}
M.~E. Peskin, and D.~V. Schroeder, \textit{An Introduction to Quantum Field Theory} (Addison-Wesley, Reading, 1995).

\end{thebibliography}
\end{document}